\documentclass[journal=jacsat,manuscript=article]{achemso}
\usepackage[version=3]{mhchem} 
\usepackage{siunitx}
\usepackage[dvipsnames]{xcolor}
\usepackage{amssymb}
\usepackage{pifont}
\usepackage{wrapfig}
\usepackage{hyperref}

\usepackage{amsmath} 

\usepackage{tabularx,colortbl}
\usepackage{capt-of}
\usepackage{adjustbox}
\usepackage[font=normalsize,labelfont={bf},font={bf}]{caption}
\usepackage{graphicx}
\usepackage{subcaption}
\usepackage{tikz}
\usepackage{caption}
\usepackage{comment}
\usepackage{array}
\usepackage{multirow}
\usepackage[labelfont=md,textfont=md]{caption}

\author{Srivathsan Badrinarayanan}
\affiliation[cheme]
{Department of Chemical Engineering, Carnegie Mellon University, 15213, USA}

\author{Chakradhar Guntuboina}
\affiliation[ece]
{Department of Electrical and Computer Engineering, Carnegie Mellon University, 15213, USA}

\author{Parisa Mollaei}
\affiliation[meche]
{Department of Mechanical Engineering, Carnegie Mellon University, 15213, USA}

\author{Amir Barati Farimani}
\email{barati@cmu.edu}
\affiliation[meche]
{Department of Mechanical Engineering, Carnegie Mellon University, 15213, USA}
\alsoaffiliation[biomed]
{Department of Biomedical Engineering, Carnegie Mellon University, 15213, USA}
\alsoaffiliation[mld]
{Machine Learning Department, Carnegie Mellon University, 15213, USA}

\title[An \textsf{achemso} demo]
{Multi-Peptide: Multimodality Leveraged Language-Graph Learning of Peptide Properties}

\begin{document}

\begin{abstract}


\noindent Peptides are essential in biological processes and therapeutics. In this study, we introduce Multi-Peptide, an innovative approach that combines transformer-based language models with Graph Neural Networks (GNNs) to predict peptide properties. We combine PeptideBERT, a transformer model tailored for peptide property prediction, with a GNN encoder to capture both sequence-based and structural features. By employing Contrastive Language-Image Pre-training (CLIP), Multi-Peptide aligns embeddings from both modalities into a shared latent space, thereby enhancing the model's predictive accuracy. Evaluations on hemolysis and nonfouling datasets demonstrate Multi-Peptide's robustness, achieving state-of-the-art 86.185\% accuracy in hemolysis prediction. This study highlights the potential of multimodal learning in bioinformatics, paving the way for accurate and reliable predictions in peptide-based research and applications. 


\end{abstract}


\section{Introduction}

Peptides, composed of distinct sequences of amino acid residues, serve as essential components in numerous biological processes and applications \cite{langel2009introduction, fennema2008fennema, voet2002fundamentals}. The unique properties of peptides, such as hemolysis and fouling behavior are critical considerations in the development of effective peptide-based therapeutics and biomaterials \cite{dunn2015peptide}. Hemolysis, characterized by the disruption of red blood cells, is particularly significant in the design of peptide-based drugs \cite{hemolysis}. Peptides with nonfouling attributes demonstrate reduced interactions with surrounding molecules, making them highly desirable for diverse biomedical applications \cite{nonfouling}. 
The structural and functional attributes of peptides, shaped by their specific arrangement of amino acids and overall length, govern how they interact with biological targets and their surroundings \cite{petsko2004protein, schulz2013principles}. Understanding this complex interplay is essential for designing customized biomaterials and developing therapeutic strategies\cite{DEGRADO198851, varanko2020recent, FOSGERAU2015122}. Traditionally, computational methodologies including quantitative structure-activity relationship (QSAR) models \cite{QSARreview}, have established links between peptide sequences and specific structural properties. However, such methods face challenges in scalability and computational efficiency, particularly when dealing with exponentially expanding sequence repositories \cite{Golbraikh2016}. This underscores the urgent need for innovative computational approaches to decipher the intricate properties associated with protein sequences. In recent years, machine learning techniques have significantly transformed these methodologies and provide advanced capabilities for data analysis and prediction\cite{yadav2022prediction, mollaei2023activity, kim2024gpcr, mollaei2023unveiling, mollaei2024idp}. Machine learning models are uniquely suited to leverage the vast amounts of biological data. The exponential growth in biological data, including protein sequence data through continuous additions to the Protein Data Bank \cite{proteindatabank, 10.1093/nar/28.1.235}, propelled by advancements in high-throughput sequencing technologies, presents lots of opportunities for predictive modeling of protein properties. In addition to the growth of the protein sequence databases, the advent of advanced protein structure modeling systems such as Google DeepMind's AlphaFold \cite{jumper2021highly}, has significantly accelerated efforts to understand the relationship between protein structure and its functional properties. AlphaFold, which leverages deep learning and AI techniques to predict a protein's intricate 3D architecture from its amino acid sequence, has enabled researchers to bridge the longstanding knowledge gap between sequence and structure in molecular biology \cite{10.1093/nar/gkab1061}. The widespread availability of accurate protein structure predictions now enables researchers to integrate comprehensive structural information into predictive modeling, moving beyond the constraints of relying solely on sequence information.

\begin{figure}[t!]
     \centering
     \includegraphics[width=0.95\linewidth]{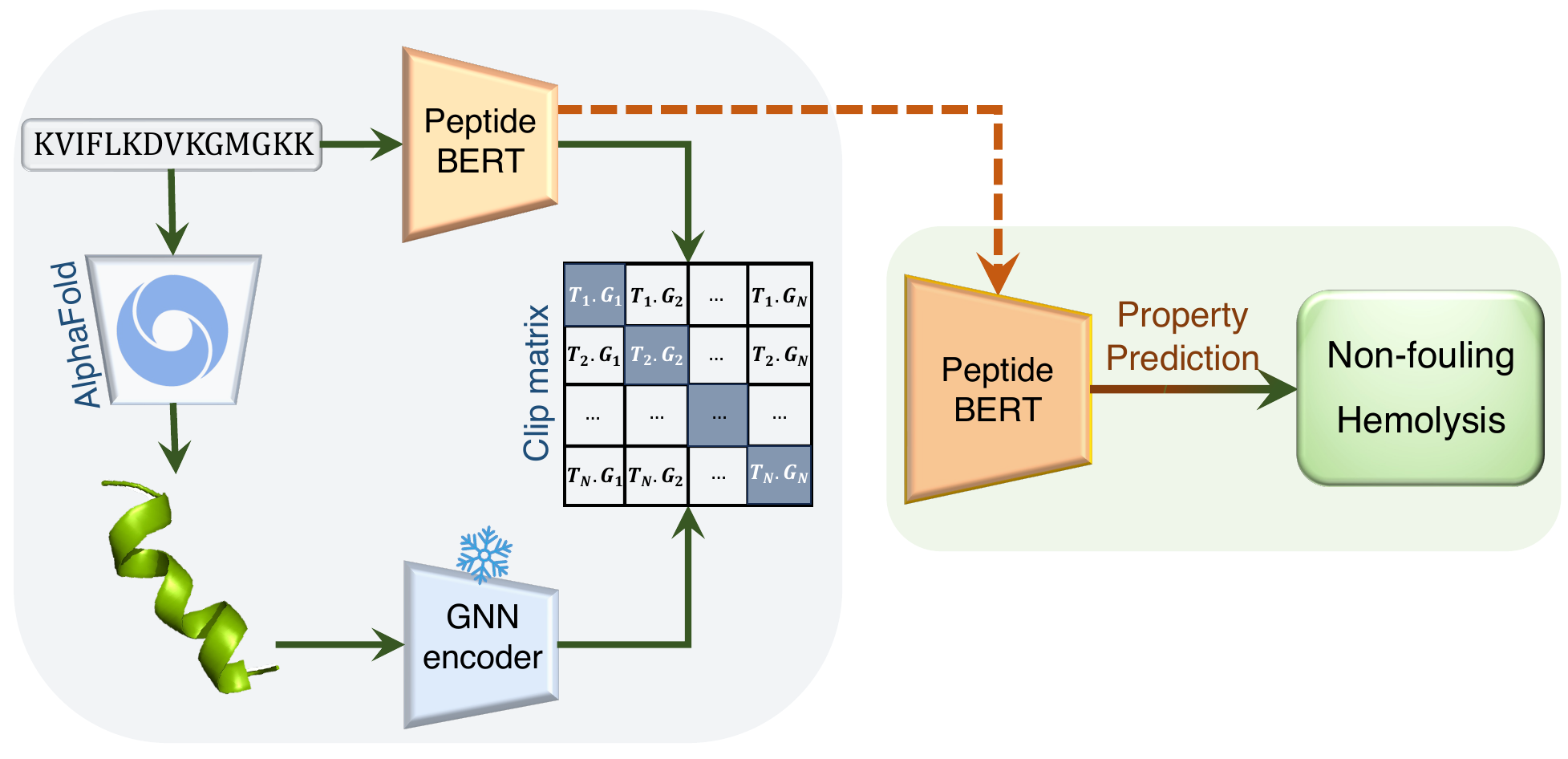}
     \caption{Representation of the Multi-Peptide framework. This figure shows the pre-training of the PeptideBERT and GNN encoders, showcasing the CLIP's ability to further train the BERT model by freezing the GNN weights. Inference is done at the end using the updated PeptideBERT weights on each of the test datasets.}
     \label{fig:framework}
 \end{figure}

Deep learning, particularly inspired by transformers and large language models (LLMs), has heralded a transformative era in protein structure prediction, providing data-driven insights into the interactions among constitutent amino acids \cite{doi:10.1021/acscatal.3c02743, 10.1093/bib/bbad358, ferruz2022deep}. While previous work like PeptideBERT \cite{peptidebert} excel at understanding protein properties based on just their sequences, they may lack the ability to directly incorporate spatial arrangements and interactions among amino acids within the protein structure. The incorporation of another modality, such as protein structural information improves the model's understanding and thereby enhances the predictive modeling capacity. By leveraging Graph Neural Networks (GNNs) \cite{4700287} to encode the three-dimensional structure of peptides, we can capture local interactions, spatial arrangements, and other structural features that are not explicitly represented in the sequence data alone.

In this study, we introduce Multi-Peptide, an innovative multimodality leveraged language-graph learning approach for peptide properties. Multi-Peptide combines PeptideBERT \cite{peptidebert}, a transformer-based language model fine-tuned for peptide property prediction, with a Graph Neural Network (GNN) encoder to learn complex representations of amino acid sequences. We use protein sequences from datasets corresponding to hemolysis and nonfouling behavior \cite{doi:10.1021/acs.jcim.2c01317}, and generate the Protein Data Bank (PDB) files corresponding to each protein sequence using AlphaFold. By pre-training the PeptideBERT transformer model individually on the sequence data and the GNN model individually on the PDB graph data, we take advantage of making each model learn before the final downstream prediction. After initial pre-training, we employ a variant of Contrastive Language Image Pre-training (CLIP) \cite{clip} in our ensemble, to enable the PeptideBERT model to learn better, by synergistically combining the global contextual understanding of PeptideBERT with the GNN's capability to capture local sequence patterns. By leveraging the latent space alignment of the individual model embeddings, this approach represents a promising advancement in peptide property prediction, offering a new methodology for the property prediction of peptides based on both their sequence and structure.

\section{Methods}

\subsection{Datasets}

\begin{figure}[t!]
     \centering
     \includegraphics[width=0.95\linewidth]{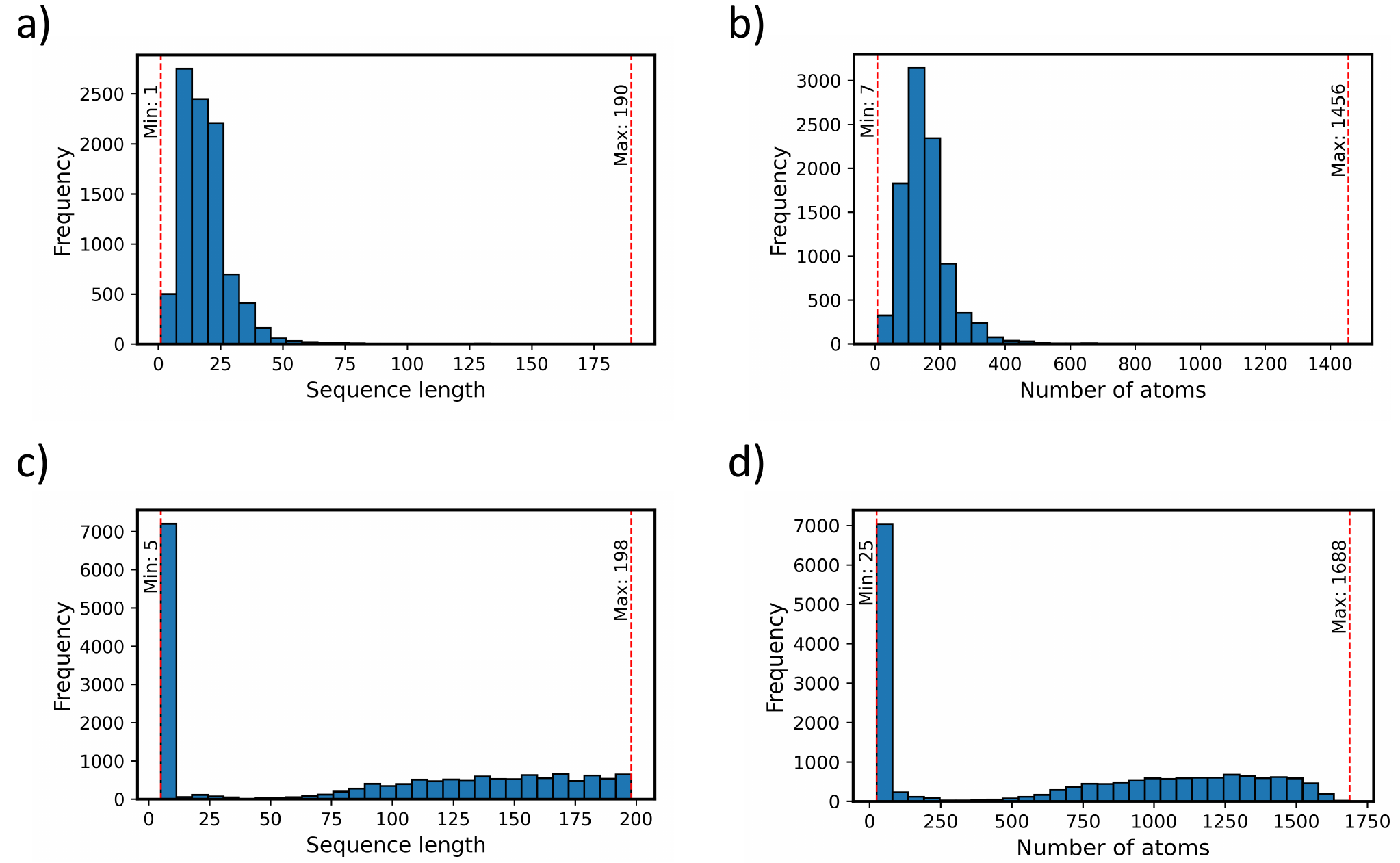}
     \caption{Distribution of peptide sequence lengths across the hemolysis and non-fouling datasets. (a) Sequence lengths vary from 1 to 190 amino acids in the hemolysis dataset. The distribution is not uniform, showing a prominent peak and a spread. (b) Number of atoms vary from 7 to 1456 in the hemolysis dataset. The distribution mirrors the corresponding sequence length distribution. (c) Sequence lengths range from 5 to 198 amino acids in the nonfouling dataset. The distribution is relatively uniform at higher lengths, with a peak at lower values and sparse occurrences over a wide range. (d) Number of atoms vary from 25 to 1688 in the nonfouling dataset. The distribution mirrors the corresponding sequence length distribution.}
     \label{fig:nf_atoms}
 \end{figure}

The datasets for hemolysis and nonfouling behavior consist of letter sequences paired with labels indicating whether each sequence corresponds positively or negatively to the respective property\cite{doi:10.1021/acs.jcim.2c01317}. The sequences are of various lengths as seen in figure \ref{fig:nf_atoms}.  For this study, the sequence data from the referenced dataset is fed into the AlphaFold system to gather more information about the peptide sequence. The output of AlphaFold is a Protein Data Bank (PDB) containing detailed atomic coordinates, which provides crucial insights into the structural arrangement of proteins.

Computational techniques are employed to predict hemolytic properties, utilizing the Database of Antimicrobial Activity and Structure of Peptides (DBAASPv3) \cite{10.1111/1574-6968.12489}. Due to experimental variability, the sequences appear multiple times with different labels in the dataset. After removing duplicates with both positive and negative labels, we are left with 845 positively marked sequences (15.065\%) and 4764 negatively marked sequences (84.935\%).

Information for forecasting resistance against nonspecific interactions (nonfouling) is gathered from a study \cite{https://doi.org/10.1002/pep2.24079}, and employed on a dataset of 3,600 positively marked sequences and a dataset of 13,585 negatively marked sequences. Here, positively marked indicates a nonfouling protein sequence. Removing 7 sequences which were duplicates having the same label, and 3 sequences for which AlphaFold failed to generate the corresponding structural PDB files, we are left with 3596 positively marked sequences (20.937\%)  and 13579 negatively marked sequences (79.063\%).  Negative examples are derived from insoluble and hemolytic peptides, along with scrambled negatives (with a length similar to the positive sequences), following an approach outlined in a referenced work \cite{C2SC21135A}.


The datasets were preprocessed via a custom encoding method, where each of the 20 amino acids was represented by its corresponding index in a predefined array. To ensure compatibility with our ensemble, we needed to ensure that the data could be used by both the PeptideBERT \cite{peptidebert} model and the GNN. For PeptideBERT, which is built on top of ProtBERT \cite{protBERT}, the datasets were converted back from integers to letter character using reverse mapping and then re-encoded using ProtBERT's encoding scheme. For the Graph Neural Network (GNN), the input data consisted of features extracted from AlphaFold-generated PDB files. These features included atom coordinates (x, y, z), atomic number, atomic mass, atomic radius, indication of whether the atom is part of a sidechain or backbone, residue index, number of atoms in the residue, and residue sequence number.


The available datasets are imbalanced, with a higher number of negative examples compared to positive examples. To address this challenge, we employed a balancing technique known as oversampling, where we increase the number of positive examples by duplicating them. This technique ensures that the model is trained on a balanced dataset, thereby preventing the model from being biased towards the majority class. We finally split each dataset into two non-overlapping subsets: a training set (to train the model) consisting of 80\% of the entire data set, and a test set (to benchmark the model’s generalization performance on unseen data) consisting of the remaining 20\% of the data set.

\subsection{Model Architecture}

In this study, we propose an innovative approach that aims to improve the prediction accuracy of a transformer-based language model for the specific peptide properties by introducing training in an additional modality. Building upon previous work, PeptideBERT \cite{peptidebert}, which was fine-tuned over the pre-trained ProtBERT \cite{protBERT} transformer model to predict the peptide properties by taking just the protein sequences as inputs, we enhanced the predictive capabilities of the existing framework by introducing another modality of data (in this case, graphical data), to capture additional dependencies. The transformer-based BERT model \cite{BERT}, which has the attention mechanism at its heart \cite{attention}, captures long-range dependencies and global context. The additional modality, in this case the graph-based learning through Graph Neural Networks over the protein structure data, augments the transformer model's understanding by focusing on local features and structural relationships. 

The model architecture comprises three key components: a Graph Neural Network (GNN), the pretrained language model (PeptideBERT), and the projection heads for projecting embeddings into the shared latent space for the CLIP loss matrix computation across the two modalities. The 11 features corresponding to each atom (extracted from the PDB files) are the nodes of the GNN, and the edges represent the relationship between the nodes. The GNN module, leveraging PyTorch Geometric's \cite{DBLP:journals/corr/abs-1903-02428} SAGEConv layer, conducts graph convolution on protein sequence graphs, aggregating information from neighboring nodes iteratively. It is followed by a fully connected neural network incorporating ReLU activation functions and a sigmoid layer, for the extraction of graph embeddings. On the other hand, PeptideBERT processes protein sequences and their corresponding attention masks to generate contextual text embeddings through ProtBERT. Additionally, projection heads are used to map the graph and text embeddings into a unified latent space, to implement contrastive learning through CLIP loss. These projection heads consist of linear projection layers, GELU activation, fully connected layers, dropout, and layer normalization. The integration of the GNN, PeptideBERT, and projection heads to project embeddings onto a shared latent space within a single module facilitates the joint learning of structural and textual representations from protein sequences.

Our approach leverages pre-training \cite{thrun1998learning} \cite{pan2009survey} individually for both the PeptideBERT model and the GNN model, which will later be used in the ensemble as above. By leveraging pre-trained weights for each model, acquired from training over each of the datasets, our models gain the ability to generalize their knowledge to specific tasks. This process of knowledge transfer enables each model to summarize high-level features and correlations relevant to peptide and protein properties, leading to better performance and faster convergence during training. The pre-trained individual models are then combined together in the same latent space, using a variant of the Contrastive Language-Image Pretraining (CLIP) loss \cite{clip} to then undergo fine-tuning on a targeted task of peptide protein property prediction. 

CLIP, pioneered by OpenAI, extends the capabilities of pre-trained individual models by enabling them to understand and reason across different modalities. CLIP operates on the principle of contrastive learning, where the model is trained to associate similar pairs of inputs while distinguishing dissimilar pairs. At its core, the CLIP loss function employs a softmax operation to normalize similarity scores between GNN and PeptideBERT embeddings, modulated by a temperature parameter. These normalized similarity scores serve as targets for the model's predictions, ensuring that the model learns to align embeddings from both modalities in a manner that reflects their semantic and structural correspondence. 

The loss computation involves calculating cross-entropy loss between predicted logits, derived from the dot product of PeptideBERT embeddings and transposed GNN embeddings, and the target similarity scores. It is important to note that CLIP is performed on the embeddings from the baseline ProtBERT model of PeptideBERT, since PeptideBERT is a fine-tuned model built over ProtBERT. By optimizing over the loss function, the model learns to associate relevant textual descriptions with corresponding graphical information and vice versa, enabling accurate classification and retrieval tasks across diverse inputs. 

In the context of our study, CLIP loss encourages the model to learn meaningful representations by contrasting peptide sequences with their associated binary protein properties. For a particular protein, say \( p \), the graph (g) and text (t) embeddings generated by the GNN and PeptideBERT encoders are represented by
\[ e_g = \text{GNN}(\text{structure}(p)) \quad \text{and} \quad e_t = \text{PeptideBERT}(\text{sequence}(p)), \]
respectively. Let the similarity between two vectors (in this case the two modalities' embeddings) \( x \) and \( y \) be measured and represented through the function \( \text{sim}(x, y) \), where the function 
\( \text{sim}(x, y) \) is the dot product between the normalized embeddings, reflecting the cosine similarity. The temperature parameter of the analysis is denoted by \( \tau \), and the overall symmetric loss between the graph and text modalities (denoted by g and t respectively), \( L(\text{g}, \text{t}) \) is given by
\[ L(\text{g}, \text{t}) = \frac{1}{2} \left[ l(\text{g}, \text{t}) + l(\text{t}, \text{g}) \right], \]
where
\[ l(\text{g}, \text{t}) = -\sum_{i=1}^{N} \log \left( \frac{e^{\text{sim}(e_g^i, e_t^i) / \tau}}{\sum_{j=1}^{N} e^{\text{sim}(e_g^i, e_t^j) / \tau}} \right). \]

In the CLIP algorithm, the embeddings \(e_g \) and \(e_t\)  are normalized, and the similarities are computed as dot products. The targets are computed as a softmax over the average of the similarities from both modalities. This results in the above loss, and similarly for \(l\)(t,g). 

The objective of the CLIP loss in this binary classification task is to train a model to accurately discriminate between two classes (say, positive and negative) by learning representations that effectively capture the semantic and graphical content of the data. Through a contrastive learning framework, CLIP aims to map textual descriptions (in this case, protein sequences) and corresponding graphs (protein structure information) to a shared embedding space where similar instances are positioned close together while dissimilar instances are pushed farther apart. The CLIP loss encourages the model to assign distinct and discriminative representations to positive and negative examples, facilitating better separation between the two classes. 

In our study, we backpropagate through the CLIP loss matrix after pre-training each model (PeptideBERT and GNN) individually over each protein property dataset. We freeze the pre-trained GNN model's weights during backpropagation, allowing the algorithm to update the PeptideBERT model's weights. After the weights are updated and tuned, the PeptideBERT model's weights are used for inference on unseen protein sequences, predicting a latent space embedding vector for each sequence. By comparing these latent vectors with those of the training samples, the most similar output is chosen as the output for the test sequence. Therefore, by leveraging pre-trained individual models to incorporate multimodality in training using CLIP loss, the PeptideBERT model gains a deeper understanding of the relationships between peptide sequences and protein properties.

\section{Results and Discussion}

The PeptideBERT model and the GNN model were each trained individually on both of the datasets. The specific parameters in each of the model architectures, and the relevant hyperparameters used during pre-training are shown in \nameref{section:supportinginfo} under the section called \nameref{section:architectureandparams}. 

The individually pre-trained models were then projected onto the same latent space with the CLIP loss implemented. The CLIP ensemble was trained for 100 epochs with a batch size of 20.  For CLIP training, a learning rate of 6.0e-5 was employed. The learning rate scheduler utilized was the Learning Rate on Plateau, used to reduce the learning rate by a factor of 0.4 for a patience of 5 epochs. These parameter settings were meticulously chosen and fine-tuned through iterative experimentation to maximize the model's effectiveness in the targeted research task. For each task, a separate model was trained on the corresponding dataset. The models were trained using the AdamW optimizer, and the binary cross-entropy loss function was employed within the larger CLIP loss. Training was performed on a single NVIDIA GeForce RTX 2080Ti GPU with 4 cores and 11GiB of memory in each core. 

Once the ensemble is trained, the weights from the CLIP matrix are extracted. The weights corresponding to the BERT transformer model are then used for inference on each of the test datasets. We expect the transformer model to have learned synergistically with the GNN model, capturing dependencies based on the features extracted and correlations learned in coherence. As seen in table \ref{accuracytable}, the accuracy of the multimodality-leveraged transformer being on par with a fine-tuned PeptideBERT and other models \cite{peptidebert}\cite{timmons2020happenn}, demonstrated the advantages of the introducing and leveraging an additional mode of data. It is important to note that we have benchmarked the Multi-Peptide model against the individual PeptideBERT and GNN components as well. Multi-Peptide's BERT gives state-of-the-art (SOTA) results for the Hemolysis dataset in particular. The improvement in Multi-Peptide's accuracy for the Hemolysis dataset even though the individual components have lower accuracy demonstrates the capability of the CLIP-implemented ensemble to learn from the presence of multiple modalities.

\begin{table}[h]
\centering
\caption{Accuracies of models on different datasets}
\label{accuracytable}
\begin{tabular}{ |c|c|c| }
 \hline
 Dataset & Model & Accuracy (\%) \\
 \hline
 Hemolysis & Multi-Peptide's BERT (this study) & \textbf{86.185}\\ 
  & Fine-tuned PeptideBERT \cite{peptidebert} & 86.051  \\ 
  & GNN & 84.94 \\
  & HAPPENN \cite{timmons2020happenn} & 85.7 \\
  \hline
 Non-fouling & Multi-Peptide's BERT (this study) &  82.431 \\ 
  & Fine-tuned PeptideBERT \cite{peptidebert} & \textbf{88.365} \\ 
  & GNN & 79.05 \\
  & Embedding + LSTM \cite{peptidebert} & 82.0 \\
 \hline
\end{tabular}
\end{table}

In this particular analysis, however, the accuracy of Multi-Peptide's ensemble does not surpass that of the fine-tuned PeptideBERT for the non-fouling dataset. This may be due to several factors. Introducing a GNN model into the ensemble significantly increases overall complexity, necessitating more extensive and precise data for effective training compared to the sequence-based transformer model. The effectiveness of the GNN relies heavily on the AlphaFold model's capability to generate highly accurate protein structure data. However, protein structure data can be noisy \cite{ctx9980881200004436} or less directly predictive of the binary property compared to sequence data. Incorporating noisy data can also degrade overall model performance. Additionally, combining sequence-based features from the transformer with structure-based features from the GNN presents challenges in effectively aligning these different feature types. If the features are not well-aligned or complementary, the ensemble struggles to learn useful representations.

Moreover, it is crucial to consider that PeptideBERT \cite{peptidebert} was trained on a heavily imbalanced and augmented dataset. In contrast, Multi-Peptide was trained on a balanced dataset achieved through oversampling and relied solely on data from primary sources without additional augmentations. While PeptideBERT excels in the specific property prediction task due to its fine-tuning, our study highlights the robustness and capability of Multi-Peptide to predict protein properties by integrating both sequence and structural data. Here, it is also important to note that hyperparameter tuning was not conducted on each dataset, but was rather done to improve the overall performance of the Multi-Peptide model on the whole. This demonstrates the potential of Multi-Peptide to provide a more holistic understanding of protein properties, even though its current performance accuracy on the nonfouling dataset is slightly lower compared to the specialized fine-tuned PeptideBERT.

\subsection{Visualization of Representations}

\begin{figure}[t!]
     \centering
     \includegraphics[width=0.95\linewidth]{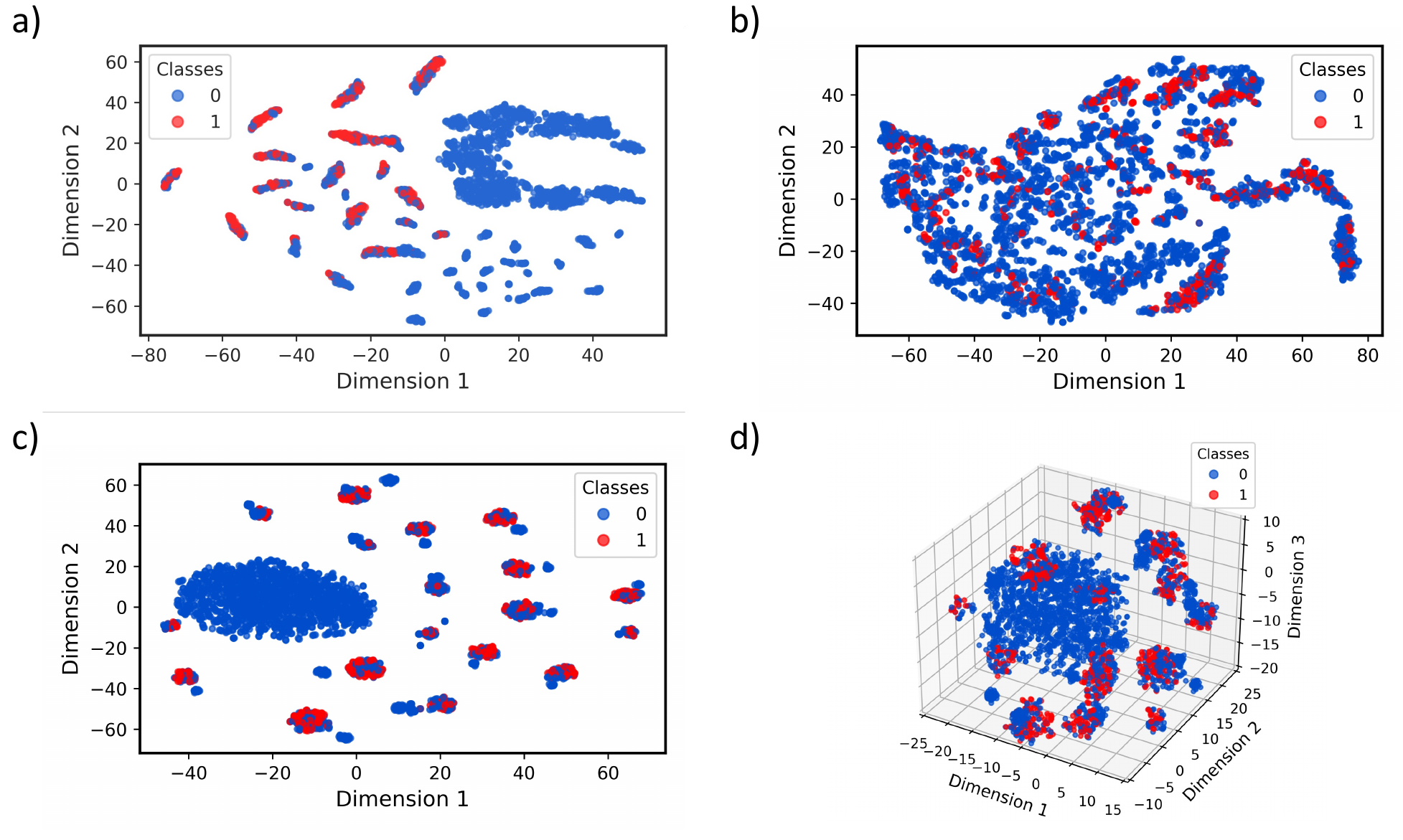}
     \caption{tSNE plots for the embeddings corresponding to the nonfouling test dataset. The blue points represent the negatively marked sequences, while the red points denote the positives. (a) 2D tSNE plot of the embedding space generated by the PeptideBERT encoder. (b) 2D tSNE plot of the embedding space generated by the GNN encoder. (c) 2D tSNE plot for the shared latent space after CLIP. (d) 3D tSNE plot for the shared latent space after CLIP.}
     \label{fig:tSNE}
 \end{figure}

To further explain the performance and capabilities of our models, we employ t-distributed Stochastic Neighbor Embedding (t-SNE) to visualize the embedding spaces derived from various methodologies. t-SNE \cite{JMLR:v9:vandermaaten08a} is a powerful tool for dimensionality reduction that helps in understanding the structure of high-dimensional data by projecting it into a lower-dimensional space. By visualizing these embeddings as seen in figure \ref{fig:tSNE}, we aim to provide a qualitative analysis of how well each model captures the underlying patterns and relationships within the peptide sequences. This visualization will help us understand the clustering behavior and the degree of separation between different classes, which are crucial for the accuracy and effectiveness of the models. In this study, we utilized datasets with a significant class imbalance, with a notably larger number of negatively marked sequences compared to positively marked ones. The t-SNE visualizations for PeptideBERT, GNN, and post-CLIP embeddings offer critical insights into how each model handles this imbalance and achieves class separation. In particular, this section aims to qualitatively explain the CLIP embedding process for the non-fouling dataset to understand if CLIP enhances the separation of classes, even though the test accuracy seems to be slightly lower than that of the fine-tuned PeptideBERT model.

The t-SNE plot of PeptideBERT embeddings revealed a central cluster of negatively marked sequences, surrounded by smaller, dispersed groups containing both negatively and positively marked sequences. This indicated that the PeptideBERT model captures semantic similarities within each class but also shows overlap between classes, highlighting the challenge of clear class separation with the BERT architecture alone.

Contrarily, the t-SNE plot of GNN embeddings demonstrated a more segregated distribution. A significant pattern of negatively marked sequences occupied the majority of the plot, with positives mirroring a similar pattern, though with some segregation. This segregation highlighted the GNN’s potential to encode structural relationships within peptide sequences, leading to potential class-specific clusters after CLIP.

Following Contrastive Language-Image Pre-training (CLIP), the t-SNE plots showed a prominent central cluster of negatively marked sequences, surrounded by smaller, more concentrated patches of mixed classes. These mixed patches appeared closer smaller and more tight knit compared to the PeptideBERT embeddings, suggesting that CLIP enhanced the discrimination between classes in the reduced-dimensional space. This is more prominent in the 3D t-SNE plot following the CLIP implementation, where we can observe some separation within the smaller mixed clusters in the 3D space as well. This improvement in class separation highlights the efficacy of CLIP in capturing distinct class patterns and mitigating the impact of class imbalance.

The t-SNE visualizations provide valuable insights into the strengths and limitations of each embedding technique for peptide sequence classification. Through the embeddings, we observe that PeptideBERT demonstrated semantic understanding, GNN captured structural relationships, and the CLIP process enabled improved class discrimination. These visualizations underscore the potential of multimodal pre-training approaches such as the one proposed in this paper.

\section{Conclusion}

In this study, we introduced Multi-Peptide, a novel approach that leverages multimodality in machine learning to enhance the prediction accuracy of peptide properties. By integrating a Graph Neural Network (GNN) with the transformer-based PeptideBERT model, we aimed to capture both the sequence-based and structural features of peptides. Our approach utilizes a variant of Contrastive Language-Image Pre-training (CLIP) to align the embeddings from these two modalities into a shared latent space, thereby facilitating more robust and comprehensive representations.

The results from our experiments demonstrate the potential of Multi-Peptide in advancing peptide property prediction. Integration of GNN and transformer-based models allows us to capture a broader range of features. Although the accuracy of Multi-Peptide was comparable to that of the fine-tuned PeptideBERT, this highlights the robustness of our approach in handling complex data structures and extracting meaningful features from both sequence and structural information. Our t-SNE visualizations, especially the post-CLIP embeddings provided valuable insights, showing enhanced class discrimination, highlighting the efficacy of multimodal pre-training in mitigating class imbalance and improving classification performance, and underscoring the strength of our approach in providing a more holistic understanding of protein properties.

In conclusion, Multi-Peptide represents a significant step forward in leveraging multimodality for peptide property prediction. Our approach holds promise for enhancing the accuracy and robustness of predictive models, offering a deeper understanding of peptide characteristics. Future work will focus on refining the integration of modalities, and further optimizing the model architecture to fully harness the complementary strengths of sequence-based and structural features. This study underscores the importance and potential of multimodal learning in advancing the field of bioinformatics, through Multi-Peptide's ability to integrate sequence and structural data.

\section{Data and Software Availability}
The necessary code and data used in this study can be accessed here:
\url{https://github.com/srivathsanb14/MultiPeptide}

\section{Supporting Information}
\label{section:supportinginfo}

\subsection{Model training}
\label{section:architectureandparams}

Table \ref{modelarchitecturetable} shows the parameters associated with the individual PeptideBERT and GNN model architectures. These models were used for individual pre-training, and the respective hyperparameters used as also listed in the same table. These hyperparameters were meticulously chosen after thorough analysis. This step of pre-training was done for 50 epochs each, with a batch size of 20. The projection dimension of the GNN was made to be 2048 to enhance model projection and thereby capture more information. The GNN had 11 input features, as listed before. The PeptideBERT configuration included a hidden size of 256, 8 hidden layers, 8 attention heads, and a dropout rate of 0.10, with a vocabulary size of 25. 

\begin{table}[h]
\centering
\begin{minipage}{.45\linewidth}
  \centering
  \begin{tabular}{|c|c|}
  \hline
  PeptideBERT Parameter & Value \\
  \hline
  Number of input parameters & 1 \\
  Vocabulary size & 25 \\
  Hidden size & 256 \\
  Hidden layers & 8 \\
  Attention heads & 8 \\
  Dropout & 0.10 \\
  Learning rate (pre-training) & 1e-5 \\
  Scheduler factor (pre-training) & 0.1 \\
  Scheduler patience (pre-training) & 6 \\
  \hline
  \end{tabular}
  \label{tab:peptidebert}
\end{minipage}%
\hspace{0.05\linewidth}
\begin{minipage}{.45\linewidth}
  \centering
  \begin{tabular}{|c|c|}
  \hline
  GNN Parameter & Value \\
  \hline
  Number of input parameters & 11 \\
  Hidden dimension & 2048 \\
  Learning rate (pre-training) & 1e-4 \\
  Scheduler factor (pre-training) & 0.4 \\
  Scheduler patience (pre-training) & 10 \\
  \hline
  \end{tabular}
  \label{tab:gnn}
\end{minipage}
\caption{PeptideBERT and GNN model architectures}
\label{modelarchitecturetable}
\end{table}

The hyperparameters associated with the CLIP process have been listed earlier. Table \ref{time_table} shows the CLIP training time for each dataset. It is important to note that this process is carried out after the individual pre-training step. The table showcases the time complexity of the CLIP process by showing the training time for each dataset.

\begin{table}[h]
\begin{tabular}{|c|c|c|c|}
\hline
Dataset & Training time (minutes) \\
\hline
Hemolysis & 59 \\
Non-fouling & 194 \\
\hline
\end{tabular}
\caption{Training time for the CLIP process for each dataset}
\label{time_table}
\end{table}

\subsection{Reproducibility and other comments}
\label{section:reproducibility}

In order to confirm the reproducibility of our results, the CLIP training process was conducted multiple times. This training process was repeated, keeping the same set of hyperparameters as earlier. Table \ref{resultstable} showcases the accuracy recorded at the inference stage after each training run, demonstrating that the model is robust and only susceptible to the stochastic nature of the training and inference processes. 

\begin{table}[h]
\begin{tabular}{|c|c|c|c|}
\hline
Dataset & Run 1 & Run 2 & Run 3\\
\hline
Hemolysis & 84.937 \% & 85.383 \% & 85.739 \% \\
Non-fouling & 82.421 \% & 81.839 \% & 82.131 \% \\
\hline
\end{tabular}
\caption{PeptideBERT and GNN model architectures}
\label{resultstable}
\end{table}

Other tests like ablation studies are not conducted as the primary objective of this study is to demonstrate the efficacy of the Multi-Peptide approach rather than to analyze the contribution of each individual component. The individual components of each model are well-studied, with the accuracy of the PeptideBERT model documented in existing literature too. Since our study entirely focuses on the synergistic understanding between the PeptideBERT and GNN models, it would be futile to mask individual components for ablation studies. 

\begin{acknowledgement}
We acknowledge the contributions of various individuals and organizations that have made this study possible. This includes the providers of the datasets used in our study, the developers of AlphaFold and PyTorch Geometric, and the teams behind ProtBERT and PeptideBERT. Additionally, we express our gratitude to our colleagues for their valuable feedback and suggestions, which have greatly improved the quality of this work. 
\end{acknowledgement}

\bibliography{reference}

\providecommand{\latin}[1]{#1}
\makeatletter
\providecommand{\doi}
  {\begingroup\let\do\@makeother\dospecials
  \catcode`\{=1 \catcode`\}=2 \doi@aux}
\providecommand{\doi@aux}[1]{\endgroup\texttt{#1}}
\makeatother
\providecommand*\mcitethebibliography{\thebibliography}
\csname @ifundefined\endcsname{endmcitethebibliography}  {\let\endmcitethebibliography\endthebibliography}{}
\begin{mcitethebibliography}{42}
\providecommand*\natexlab[1]{#1}
\providecommand*\mciteSetBstSublistMode[1]{}
\providecommand*\mciteSetBstMaxWidthForm[2]{}
\providecommand*\mciteBstWouldAddEndPuncttrue
  {\def\EndOfBibitem{\unskip.}}
\providecommand*\mciteBstWouldAddEndPunctfalse
  {\let\EndOfBibitem\relax}
\providecommand*\mciteSetBstMidEndSepPunct[3]{}
\providecommand*\mciteSetBstSublistLabelBeginEnd[3]{}
\providecommand*\EndOfBibitem{}
\mciteSetBstSublistMode{f}
\mciteSetBstMaxWidthForm{subitem}{(\alph{mcitesubitemcount})}
\mciteSetBstSublistLabelBeginEnd
  {\mcitemaxwidthsubitemform\space}
  {\relax}
  {\relax}

\bibitem[Langel \latin{et~al.}(2009)Langel, Cravatt, Graslund, Von~Heijne, Zorko, Land, and Niessen]{langel2009introduction}
Langel,~U.; Cravatt,~B.~F.; Graslund,~A.; Von~Heijne,~N.; Zorko,~M.; Land,~T.; Niessen,~S. \emph{Introduction to Peptides and Proteins}; CRC Press, 2009\relax
\mciteBstWouldAddEndPuncttrue
\mciteSetBstMidEndSepPunct{\mcitedefaultmidpunct}
{\mcitedefaultendpunct}{\mcitedefaultseppunct}\relax
\EndOfBibitem
\bibitem[Fennema \latin{et~al.}(2008)Fennema, Damodaran, and Parkin]{fennema2008fennema}
Fennema,~O.~R.; Damodaran,~S.; Parkin,~K.~L. \emph{Fennema's food chemistry}; CRC press Boca Raton, 2008\relax
\mciteBstWouldAddEndPuncttrue
\mciteSetBstMidEndSepPunct{\mcitedefaultmidpunct}
{\mcitedefaultendpunct}{\mcitedefaultseppunct}\relax
\EndOfBibitem
\bibitem[Voet \latin{et~al.}(2002)Voet, Voet, Pratt, \latin{et~al.} others]{voet2002fundamentals}
Voet,~D.; Voet,~J.~G.; Pratt,~C.~W.; others \emph{Fundamentals of biochemistry}; Wiley New York, 2002\relax
\mciteBstWouldAddEndPuncttrue
\mciteSetBstMidEndSepPunct{\mcitedefaultmidpunct}
{\mcitedefaultendpunct}{\mcitedefaultseppunct}\relax
\EndOfBibitem
\bibitem[Dunn(2015)]{dunn2015peptide}
Dunn,~B.~M. \emph{Peptide chemistry and drug design}; Wiley Online Library, 2015\relax
\mciteBstWouldAddEndPuncttrue
\mciteSetBstMidEndSepPunct{\mcitedefaultmidpunct}
{\mcitedefaultendpunct}{\mcitedefaultseppunct}\relax
\EndOfBibitem
\bibitem[Ponder(1948)]{hemolysis}
Ponder,~E. \emph{Hemolysis and Related Phenomena}; Grune \& Stratton, 1948\relax
\mciteBstWouldAddEndPuncttrue
\mciteSetBstMidEndSepPunct{\mcitedefaultmidpunct}
{\mcitedefaultendpunct}{\mcitedefaultseppunct}\relax
\EndOfBibitem
\bibitem[Yu \latin{et~al.}(2011)Yu, Zhang, Wang, Brash, and Chen]{nonfouling}
Yu,~Q.; Zhang,~Y.; Wang,~H.; Brash,~J.; Chen,~H. Anti-fouling bioactive surfaces. \emph{Acta Biomaterialia} \textbf{2011}, \emph{7}, 1550--1557\relax
\mciteBstWouldAddEndPuncttrue
\mciteSetBstMidEndSepPunct{\mcitedefaultmidpunct}
{\mcitedefaultendpunct}{\mcitedefaultseppunct}\relax
\EndOfBibitem
\bibitem[Petsko and Ringe(2004)Petsko, and Ringe]{petsko2004protein}
Petsko,~G.~A.; Ringe,~D. \emph{Protein structure and function}; New Science Press, 2004\relax
\mciteBstWouldAddEndPuncttrue
\mciteSetBstMidEndSepPunct{\mcitedefaultmidpunct}
{\mcitedefaultendpunct}{\mcitedefaultseppunct}\relax
\EndOfBibitem
\bibitem[Schulz and Schirmer(2013)Schulz, and Schirmer]{schulz2013principles}
Schulz,~G.~E.; Schirmer,~R.~H. \emph{Principles of protein structure}; Springer Science \& Business Media, 2013\relax
\mciteBstWouldAddEndPuncttrue
\mciteSetBstMidEndSepPunct{\mcitedefaultmidpunct}
{\mcitedefaultendpunct}{\mcitedefaultseppunct}\relax
\EndOfBibitem
\bibitem[Degrado(1988)]{DEGRADO198851}
Degrado,~W.~F. In \emph{Design of Peptides and Proteins}; Anfinsen,~C., Edsall,~J.~T., Richards,~F.~M., Eisenberg,~D.~S., Eds.; Advances in Protein Chemistry; Academic Press, 1988; Vol.~39; pp 51--124\relax
\mciteBstWouldAddEndPuncttrue
\mciteSetBstMidEndSepPunct{\mcitedefaultmidpunct}
{\mcitedefaultendpunct}{\mcitedefaultseppunct}\relax
\EndOfBibitem
\bibitem[Varanko \latin{et~al.}(2020)Varanko, Saha, and Chilkoti]{varanko2020recent}
Varanko,~A.; Saha,~S.; Chilkoti,~A. Recent trends in protein and peptide-based biomaterials for advanced drug delivery. \emph{Advanced drug delivery reviews} \textbf{2020}, \emph{156}, 133--187\relax
\mciteBstWouldAddEndPuncttrue
\mciteSetBstMidEndSepPunct{\mcitedefaultmidpunct}
{\mcitedefaultendpunct}{\mcitedefaultseppunct}\relax
\EndOfBibitem
\bibitem[Fosgerau and Hoffmann(2015)Fosgerau, and Hoffmann]{FOSGERAU2015122}
Fosgerau,~K.; Hoffmann,~T. Peptide therapeutics: current status and future directions. \emph{Drug Discovery Today} \textbf{2015}, \emph{20}, 122--128\relax
\mciteBstWouldAddEndPuncttrue
\mciteSetBstMidEndSepPunct{\mcitedefaultmidpunct}
{\mcitedefaultendpunct}{\mcitedefaultseppunct}\relax
\EndOfBibitem
\bibitem[Cherkasov \latin{et~al.}(2014)Cherkasov, Muratov, Fourches, Varnek, Baskin, Cronin, Dearden, Gramatica, Martin, Todeschini, Consonni, Kuz’min, Cramer, Benigni, Yang, Rathman, Terfloth, Gasteiger, Richard, and Tropsha]{QSARreview}
Cherkasov,~A. \latin{et~al.}  QSAR Modeling: Where Have You Been? Where Are You Going To? \emph{Journal of Medicinal Chemistry} \textbf{2014}, \emph{57}, 4977--5010, PMID: 24351051\relax
\mciteBstWouldAddEndPuncttrue
\mciteSetBstMidEndSepPunct{\mcitedefaultmidpunct}
{\mcitedefaultendpunct}{\mcitedefaultseppunct}\relax
\EndOfBibitem
\bibitem[Golbraikh \latin{et~al.}(2016)Golbraikh, Wang, Zhu, and Tropsha]{Golbraikh2016}
Golbraikh,~A.; Wang,~X.~S.; Zhu,~H.; Tropsha,~A. In \emph{Handbook of Computational Chemistry}; Leszczynski,~J., Ed.; Springer Netherlands: Dordrecht, 2016; pp 1--48\relax
\mciteBstWouldAddEndPuncttrue
\mciteSetBstMidEndSepPunct{\mcitedefaultmidpunct}
{\mcitedefaultendpunct}{\mcitedefaultseppunct}\relax
\EndOfBibitem
\bibitem[Yadav \latin{et~al.}(2022)Yadav, Mollaei, Cao, Wang, and Farimani]{yadav2022prediction}
Yadav,~P.; Mollaei,~P.; Cao,~Z.; Wang,~Y.; Farimani,~A.~B. Prediction of GPCR activity using machine learning. \emph{Computational and Structural Biotechnology Journal} \textbf{2022}, \emph{20}, 2564--2573\relax
\mciteBstWouldAddEndPuncttrue
\mciteSetBstMidEndSepPunct{\mcitedefaultmidpunct}
{\mcitedefaultendpunct}{\mcitedefaultseppunct}\relax
\EndOfBibitem
\bibitem[Mollaei and Barati~Farimani(2023)Mollaei, and Barati~Farimani]{mollaei2023activity}
Mollaei,~P.; Barati~Farimani,~A. Activity Map and Transition Pathways of G Protein-Coupled Receptor Revealed by Machine Learning. \emph{Journal of Chemical Information and Modeling} \textbf{2023}, \emph{63}, 2296--2304\relax
\mciteBstWouldAddEndPuncttrue
\mciteSetBstMidEndSepPunct{\mcitedefaultmidpunct}
{\mcitedefaultendpunct}{\mcitedefaultseppunct}\relax
\EndOfBibitem
\bibitem[Kim \latin{et~al.}(2024)Kim, Mollaei, Antony, Magar, and Barati~Farimani]{kim2024gpcr}
Kim,~S.; Mollaei,~P.; Antony,~A.; Magar,~R.; Barati~Farimani,~A. GPCR-BERT: Interpreting Sequential Design of G Protein-Coupled Receptors Using Protein Language Models. \emph{Journal of Chemical Information and Modeling} \textbf{2024}, \emph{64}, 1134--1144\relax
\mciteBstWouldAddEndPuncttrue
\mciteSetBstMidEndSepPunct{\mcitedefaultmidpunct}
{\mcitedefaultendpunct}{\mcitedefaultseppunct}\relax
\EndOfBibitem
\bibitem[Mollaei and Barati~Farimani(2023)Mollaei, and Barati~Farimani]{mollaei2023unveiling}
Mollaei,~P.; Barati~Farimani,~A. Unveiling Switching Function of Amino Acids in Proteins Using a Machine Learning Approach. \emph{Journal of Chemical Theory and Computation} \textbf{2023}, \emph{19}, 8472--8480\relax
\mciteBstWouldAddEndPuncttrue
\mciteSetBstMidEndSepPunct{\mcitedefaultmidpunct}
{\mcitedefaultendpunct}{\mcitedefaultseppunct}\relax
\EndOfBibitem
\bibitem[Mollaei \latin{et~al.}(2024)Mollaei, Guntuboina, Sadasivam, and Farimani]{mollaei2024idp}
Mollaei,~P.; Guntuboina,~C.; Sadasivam,~D.; Farimani,~A.~B. IDP-Bert: Predicting Properties of Intrinsically Disordered Proteins (IDP) Using Large Language Models. \emph{arXiv preprint arXiv:2403.19762} \textbf{2024}, \relax
\mciteBstWouldAddEndPunctfalse
\mciteSetBstMidEndSepPunct{\mcitedefaultmidpunct}
{}{\mcitedefaultseppunct}\relax
\EndOfBibitem
\bibitem[BERNSTEIN \latin{et~al.}(1977)BERNSTEIN, KOETZLE, WILLIAMS, MEYER~Jr, BRICE, RODGERS, KENNARD, SHIMANOUCHI, and TASUMI]{proteindatabank}
BERNSTEIN,~F.~C.; KOETZLE,~T.~F.; WILLIAMS,~G. J.~B.; MEYER~Jr,~E.~F.; BRICE,~M.~D.; RODGERS,~J.~R.; KENNARD,~O.; SHIMANOUCHI,~T.; TASUMI,~M. The Protein Data Bank. \emph{European Journal of Biochemistry} \textbf{1977}, \emph{80}, 319--324\relax
\mciteBstWouldAddEndPuncttrue
\mciteSetBstMidEndSepPunct{\mcitedefaultmidpunct}
{\mcitedefaultendpunct}{\mcitedefaultseppunct}\relax
\EndOfBibitem
\bibitem[Berman \latin{et~al.}(2000)Berman, Westbrook, Feng, Gilliland, Bhat, Weissig, Shindyalov, and Bourne]{10.1093/nar/28.1.235}
Berman,~H.~M.; Westbrook,~J.; Feng,~Z.; Gilliland,~G.; Bhat,~T.~N.; Weissig,~H.; Shindyalov,~I.~N.; Bourne,~P.~E. {The Protein Data Bank}. \emph{Nucleic Acids Research} \textbf{2000}, \emph{28}, 235--242\relax
\mciteBstWouldAddEndPuncttrue
\mciteSetBstMidEndSepPunct{\mcitedefaultmidpunct}
{\mcitedefaultendpunct}{\mcitedefaultseppunct}\relax
\EndOfBibitem
\bibitem[Jumper \latin{et~al.}(2021)Jumper, Evans, Pritzel, Green, Figurnov, Ronneberger, Tunyasuvunakool, Bates, {\v{Z}}{\'\i}dek, Potapenko, \latin{et~al.} others]{jumper2021highly}
Jumper,~J.; Evans,~R.; Pritzel,~A.; Green,~T.; Figurnov,~M.; Ronneberger,~O.; Tunyasuvunakool,~K.; Bates,~R.; {\v{Z}}{\'\i}dek,~A.; Potapenko,~A.; others Highly accurate protein structure prediction with AlphaFold. \emph{Nature} \textbf{2021}, \emph{596}, 583--589\relax
\mciteBstWouldAddEndPuncttrue
\mciteSetBstMidEndSepPunct{\mcitedefaultmidpunct}
{\mcitedefaultendpunct}{\mcitedefaultseppunct}\relax
\EndOfBibitem
\bibitem[Varadi \latin{et~al.}(2021)Varadi, Anyango, Deshpande, Nair, Natassia, Yordanova, Yuan, Stroe, Wood, Laydon, Žídek, Green, Tunyasuvunakool, Petersen, Jumper, Clancy, Green, Vora, Lutfi, Figurnov, Cowie, Hobbs, Kohli, Kleywegt, Birney, Hassabis, and Velankar]{10.1093/nar/gkab1061}
Varadi,~M. \latin{et~al.}  {AlphaFold Protein Structure Database: massively expanding the structural coverage of protein-sequence space with high-accuracy models}. \emph{Nucleic Acids Research} \textbf{2021}, \emph{50}, D439--D444\relax
\mciteBstWouldAddEndPuncttrue
\mciteSetBstMidEndSepPunct{\mcitedefaultmidpunct}
{\mcitedefaultendpunct}{\mcitedefaultseppunct}\relax
\EndOfBibitem
\bibitem[Kouba \latin{et~al.}(2023)Kouba, Kohout, Haddadi, Bushuiev, Samusevich, Sedlar, Damborsky, Pluskal, Sivic, and Mazurenko]{doi:10.1021/acscatal.3c02743}
Kouba,~P.; Kohout,~P.; Haddadi,~F.; Bushuiev,~A.; Samusevich,~R.; Sedlar,~J.; Damborsky,~J.; Pluskal,~T.; Sivic,~J.; Mazurenko,~S. Machine Learning-Guided Protein Engineering. \emph{ACS Catalysis} \textbf{2023}, \emph{13}, 13863--13895\relax
\mciteBstWouldAddEndPuncttrue
\mciteSetBstMidEndSepPunct{\mcitedefaultmidpunct}
{\mcitedefaultendpunct}{\mcitedefaultseppunct}\relax
\EndOfBibitem
\bibitem[Mardikoraem \latin{et~al.}(2023)Mardikoraem, Wang, Pascual, and Woldring]{10.1093/bib/bbad358}
Mardikoraem,~M.; Wang,~Z.; Pascual,~N.; Woldring,~D. {Generative models for protein sequence modeling: recent advances and future directions}. \emph{Briefings in Bioinformatics} \textbf{2023}, \emph{24}, bbad358\relax
\mciteBstWouldAddEndPuncttrue
\mciteSetBstMidEndSepPunct{\mcitedefaultmidpunct}
{\mcitedefaultendpunct}{\mcitedefaultseppunct}\relax
\EndOfBibitem
\bibitem[Ferruz \latin{et~al.}(2022)Ferruz, Schmidt, and H{\"o}cker]{ferruz2022deep}
Ferruz,~N.; Schmidt,~S.; H{\"o}cker,~B. A deep unsupervised language model for protein design. \emph{BioRxiv} \textbf{2022}, 2022--03\relax
\mciteBstWouldAddEndPuncttrue
\mciteSetBstMidEndSepPunct{\mcitedefaultmidpunct}
{\mcitedefaultendpunct}{\mcitedefaultseppunct}\relax
\EndOfBibitem
\bibitem[Guntuboina \latin{et~al.}(2023)Guntuboina, Das, Mollaei, Kim, and Barati~Farimani]{peptidebert}
Guntuboina,~C.; Das,~A.; Mollaei,~P.; Kim,~S.; Barati~Farimani,~A. PeptideBERT: A Language Model Based on Transformers for Peptide Property Prediction. \emph{The Journal of Physical Chemistry Letters} \textbf{2023}, \emph{14}, 10427--10434, PMID: 37956397\relax
\mciteBstWouldAddEndPuncttrue
\mciteSetBstMidEndSepPunct{\mcitedefaultmidpunct}
{\mcitedefaultendpunct}{\mcitedefaultseppunct}\relax
\EndOfBibitem
\bibitem[Scarselli \latin{et~al.}(2009)Scarselli, Gori, Tsoi, Hagenbuchner, and Monfardini]{4700287}
Scarselli,~F.; Gori,~M.; Tsoi,~A.~C.; Hagenbuchner,~M.; Monfardini,~G. The Graph Neural Network Model. \emph{IEEE Transactions on Neural Networks} \textbf{2009}, \emph{20}, 61--80\relax
\mciteBstWouldAddEndPuncttrue
\mciteSetBstMidEndSepPunct{\mcitedefaultmidpunct}
{\mcitedefaultendpunct}{\mcitedefaultseppunct}\relax
\EndOfBibitem
\bibitem[Ansari and White(2023)Ansari, and White]{doi:10.1021/acs.jcim.2c01317}
Ansari,~M.; White,~A.~D. Serverless Prediction of Peptide Properties with Recurrent Neural Networks. \emph{Journal of Chemical Information and Modeling} \textbf{2023}, \emph{63}, 2546--2553, PMID: 37010950\relax
\mciteBstWouldAddEndPuncttrue
\mciteSetBstMidEndSepPunct{\mcitedefaultmidpunct}
{\mcitedefaultendpunct}{\mcitedefaultseppunct}\relax
\EndOfBibitem
\bibitem[Radford \latin{et~al.}(2021)Radford, Kim, Hallacy, Ramesh, Goh, Agarwal, Sastry, Askell, Mishkin, Clark, Krueger, and Sutskever]{clip}
Radford,~A.; Kim,~J.~W.; Hallacy,~C.; Ramesh,~A.; Goh,~G.; Agarwal,~S.; Sastry,~G.; Askell,~A.; Mishkin,~P.; Clark,~J.; Krueger,~G.; Sutskever,~I. Learning Transferable Visual Models From Natural Language Supervision. 2021\relax
\mciteBstWouldAddEndPuncttrue
\mciteSetBstMidEndSepPunct{\mcitedefaultmidpunct}
{\mcitedefaultendpunct}{\mcitedefaultseppunct}\relax
\EndOfBibitem
\bibitem[Gogoladze \latin{et~al.}(2014)Gogoladze, Grigolava, Vishnepolsky, Chubinidze, Duroux, Lefranc, and Pirtskhalava]{10.1111/1574-6968.12489}
Gogoladze,~G.; Grigolava,~M.; Vishnepolsky,~B.; Chubinidze,~M.; Duroux,~P.; Lefranc,~M.-P.; Pirtskhalava,~M. {dbaasp: database of antimicrobial activity and structure of peptides}. \emph{FEMS Microbiology Letters} \textbf{2014}, \emph{357}, 63--68\relax
\mciteBstWouldAddEndPuncttrue
\mciteSetBstMidEndSepPunct{\mcitedefaultmidpunct}
{\mcitedefaultendpunct}{\mcitedefaultseppunct}\relax
\EndOfBibitem
\bibitem[Barrett \latin{et~al.}(2018)Barrett, Jiang, and White]{https://doi.org/10.1002/pep2.24079}
Barrett,~R.; Jiang,~S.; White,~A.~D. Classifying antimicrobial and multifunctional peptides with Bayesian network models. \emph{Peptide Science} \textbf{2018}, \emph{110}, e24079\relax
\mciteBstWouldAddEndPuncttrue
\mciteSetBstMidEndSepPunct{\mcitedefaultmidpunct}
{\mcitedefaultendpunct}{\mcitedefaultseppunct}\relax
\EndOfBibitem
\bibitem[White \latin{et~al.}(2012)White, Nowinski, Huang, Keefe, Sun, and Jiang]{C2SC21135A}
White,~A.~D.; Nowinski,~A.~K.; Huang,~W.; Keefe,~A.~J.; Sun,~F.; Jiang,~S. Decoding nonspecific interactions from nature. \emph{Chem. Sci.} \textbf{2012}, \emph{3}, 3488--3494\relax
\mciteBstWouldAddEndPuncttrue
\mciteSetBstMidEndSepPunct{\mcitedefaultmidpunct}
{\mcitedefaultendpunct}{\mcitedefaultseppunct}\relax
\EndOfBibitem
\bibitem[Elnaggar \latin{et~al.}(2021)Elnaggar, Heinzinger, Dallago, Rehawi, Wang, Jones, Gibbs, Feher, Angerer, Steinegger, Bhowmik, and Rost]{protBERT}
Elnaggar,~A.; Heinzinger,~M.; Dallago,~C.; Rehawi,~G.; Wang,~Y.; Jones,~L.; Gibbs,~T.; Feher,~T.; Angerer,~C.; Steinegger,~M.; Bhowmik,~D.; Rost,~B. ProtTrans: Towards Cracking the Language of Life{\textquoteright}s Code Through Self-Supervised Learning. \emph{bioRxiv} \textbf{2021}, \relax
\mciteBstWouldAddEndPunctfalse
\mciteSetBstMidEndSepPunct{\mcitedefaultmidpunct}
{}{\mcitedefaultseppunct}\relax
\EndOfBibitem
\bibitem[Devlin \latin{et~al.}(2018)Devlin, Chang, Lee, and Toutanova]{BERT}
Devlin,~J.; Chang,~M.; Lee,~K.; Toutanova,~K. {BERT:} Pre-training of Deep Bidirectional Transformers for Language Understanding. \emph{CoRR} \textbf{2018}, \emph{abs/1810.04805}\relax
\mciteBstWouldAddEndPuncttrue
\mciteSetBstMidEndSepPunct{\mcitedefaultmidpunct}
{\mcitedefaultendpunct}{\mcitedefaultseppunct}\relax
\EndOfBibitem
\bibitem[Vaswani \latin{et~al.}(2017)Vaswani, Shazeer, Parmar, Uszkoreit, Jones, Gomez, Kaiser, and Polosukhin]{attention}
Vaswani,~A.; Shazeer,~N.; Parmar,~N.; Uszkoreit,~J.; Jones,~L.; Gomez,~A.~N.; Kaiser,~L.; Polosukhin,~I. Attention Is All You Need. \emph{CoRR} \textbf{2017}, \emph{abs/1706.03762}\relax
\mciteBstWouldAddEndPuncttrue
\mciteSetBstMidEndSepPunct{\mcitedefaultmidpunct}
{\mcitedefaultendpunct}{\mcitedefaultseppunct}\relax
\EndOfBibitem
\bibitem[Fey and Lenssen(2019)Fey, and Lenssen]{DBLP:journals/corr/abs-1903-02428}
Fey,~M.; Lenssen,~J.~E. Fast Graph Representation Learning with PyTorch Geometric. \emph{CoRR} \textbf{2019}, \emph{abs/1903.02428}\relax
\mciteBstWouldAddEndPuncttrue
\mciteSetBstMidEndSepPunct{\mcitedefaultmidpunct}
{\mcitedefaultendpunct}{\mcitedefaultseppunct}\relax
\EndOfBibitem
\bibitem[Thrun and Pratt(1998)Thrun, and Pratt]{thrun1998learning}
Thrun,~S.; Pratt,~L. \emph{Learning to learn}; Springer, 1998; pp 3--17\relax
\mciteBstWouldAddEndPuncttrue
\mciteSetBstMidEndSepPunct{\mcitedefaultmidpunct}
{\mcitedefaultendpunct}{\mcitedefaultseppunct}\relax
\EndOfBibitem
\bibitem[Pan and Yang(2009)Pan, and Yang]{pan2009survey}
Pan,~S.~J.; Yang,~Q. A survey on transfer learning. \emph{IEEE Transactions on knowledge and data engineering} \textbf{2009}, \emph{22}, 1345--1359\relax
\mciteBstWouldAddEndPuncttrue
\mciteSetBstMidEndSepPunct{\mcitedefaultmidpunct}
{\mcitedefaultendpunct}{\mcitedefaultseppunct}\relax
\EndOfBibitem
\bibitem[Timmons and Hewage(2020)Timmons, and Hewage]{timmons2020happenn}
Timmons,~P.~B.; Hewage,~C.~M. HAPPENN is a novel tool for hemolytic activity prediction for therapeutic peptides which employs neural networks. \emph{Scientific reports} \textbf{2020}, \emph{10}, 10869\relax
\mciteBstWouldAddEndPuncttrue
\mciteSetBstMidEndSepPunct{\mcitedefaultmidpunct}
{\mcitedefaultendpunct}{\mcitedefaultseppunct}\relax
\EndOfBibitem
\bibitem[Chakravarty and Porter(2022-06)Chakravarty, and Porter]{ctx9980881200004436}
Chakravarty,~D.; Porter,~L.~L. AlphaFold2 fails to predict protein fold switching. \emph{Protein science : a publication of the Protein Society.} \textbf{2022-06}, \emph{31}\relax
\mciteBstWouldAddEndPuncttrue
\mciteSetBstMidEndSepPunct{\mcitedefaultmidpunct}
{\mcitedefaultendpunct}{\mcitedefaultseppunct}\relax
\EndOfBibitem
\bibitem[van~der Maaten and Hinton(2008)van~der Maaten, and Hinton]{JMLR:v9:vandermaaten08a}
van~der Maaten,~L.; Hinton,~G. Visualizing Data using t-SNE. \emph{Journal of Machine Learning Research} \textbf{2008}, \emph{9}, 2579--2605\relax
\mciteBstWouldAddEndPuncttrue
\mciteSetBstMidEndSepPunct{\mcitedefaultmidpunct}
{\mcitedefaultendpunct}{\mcitedefaultseppunct}\relax
\EndOfBibitem
\end{mcitethebibliography}

\end{document}